\documentclass[sigconf,nonacm]{acmart}

\usepackage{subfigure}
\usepackage{siunitx}
\usepackage{enumitem}

\usepackage{algorithm}
\usepackage[noend]{algpseudocode}
\let\ReturnInline\Return
\renewcommand{\Return}{\State\ReturnInline}
\algrenewcommand\algorithmicrequire{$\rhd$}
\algrenewcommand\algorithmicensure{$\square$}

\AtBeginDocument{%
  \providecommand\BibTeX{{%
    \normalfont B\kern-0.5em{\scshape i\kern-0.25em b}\kern-0.8em\TeX}}}

\setcopyright{acmcopyright}
\copyrightyear{2018}
\acmYear{2018}
\acmDOI{XXXXXXX.XXXXXXX}

\acmConference[Conference acronym 'XX]{Make sure to enter the correct
  conference title from your rights confirmation emai}{June 03--05,
  2018}{Woodstock, NY}

\newcommand{\ignore}[1]{}

\begin{document}

\title[GVEL: Fast Graph Loading in Edgelist and Compressed Sparse Row (CSR) formats]{GVEL: Fast Graph Loading in Edgelist and\\ Compressed Sparse Row (CSR) formats}


\author{Subhajit Sahu}
\email{subhajit.sahu@research.iiit.ac.in}
\affiliation{%
  \institution{IIIT Hyderabad}
  \streetaddress{Professor CR Rao Rd, Gachibowli}
  \city{Hyderabad}
  \state{Telangana}
  \country{India}
  \postcode{500032}
}


\settopmatter{printfolios=true}

\begin{abstract}
Efficient IO techniques are crucial in high-performance graph processing frameworks like Gunrock and Hornet, as fast graph loading can help minimize processing time and reduce system/cloud usage charges. This research study presents approaches for efficiently reading an Edgelist from a text file and converting it to a Compressed Sparse Row (CSR) representation. On a server with dual 16-core Intel Xeon Gold 6226R processors and Seagate Exos 10e2400 HDDs, our approach, which we term as \textit{GVEL}, outperforms Hornet, Gunrock, and PIGO by significant margins in CSR reading, exhibiting an average speedup of $78\times$, $112\times$, and $1.8\times$, respectively. For Edgelist reading, GVEL is $2.6\times$ faster than PIGO on average, and achieves a Edgelist read rate of $1.9$ billion edges/s. For every doubling of threads, GVEL improves performance at an average rate of $1.9\times$ and $1.7\times$ for reading Edgelist and reading CSR respectively.
\end{abstract}



\maketitle

\section{Introduction}
\label{sec:introduction}
Graphs are fundamental data structures in many applications, such as computer networks, recommendation systems, and circuit design. In recent years, a number of high performance graph processing frameworks have emerged. State-of-the-art frameworks include Gunrock \cite{wang2016gunrock}, Hornet \cite{busato2018hornet}, Ligra \cite{shun2013ligra}, and Galois \cite{nguyen2013lightweight}.\ignore{These frameworks have demonstrated efficient computation on billion-scale graphs.} Their emphasis lies in accelerating graph analytics tasks by providing high-performance kernels tailored to diverse datasets.

Unfortunately, loading graph data is a significant bottleneck in such frameworks. In fact, the cost of loading data can dominate the overall processing time --- especially as computational capabilities continue to improve\ignore{\cite{gabert2021pigo}}. Gabert and Çatalyürek \cite{gabert2021pigo} observe that\ignore{even on high-performance shared-memory graph systems running billion-scale graphs,} reading the graph from file systems, on such frameworks, takes multiple orders of magnitude longer than running the computational kernel. This slowdown not only causes a disconnect for end users and a loss of productivity for researchers\ignore{/developers}, but also increases the system/cloud usage charges. Fast loading of graphs is thus, crucial\ignore{for minimizing the time it takes to start processing and analyzing the graph data}.\ignore{This motivates us to work on efficient IO techniques. These not only improve response time, but also help lower system / cloud usage charges.}

In modern frameworks like Gunrock, loading graph data from ASCII-based file formats, specifically using the Coordinate (COO) format, is a major bottleneck. To load the graph as an Edgelist, these frameworks typically follow a sequential process of opening the input file, reading the entries one by one, and inserting them into an array. If the goal is to access the graph in the Compressed Sparse Row (CSR) format, which is often the case --- due to its storage efficiency and locality benefits, additional steps are required. These include computing the out-degrees of vertices from the Edgelist, performing prefix sum to determine the offsets of outgoing edges in the CSR representation, and then populating the CSR arrays with edges from the Edgelist. All these operations are carried out sequentially, contributing to the overall loading time.

Many graph processing frameworks\ignore{have showcased efficient computation on large-scale graphs, they}, thus, still rely on sequential I/O. This is likely due to the belief that I/O devices tend to be slow (relative to the CPU), and that achieving parallel I/O necessitates specialized systems.\ignore{Graph and matrix I/O times are seldom reported in the literature.} However, modern IO devices are fast, and implementing only sequential I/O fails to exploit the capabilities of modern Hard Disk Drives (HDDs), Redundant Array of Independent Disks (RAID) controllers, and Non-Volatile Memory (NVM) \cite{gabert2021pigo}. A number of disk-based out-of-memory graph processing systems/frameworks\ignore{\cite{zhu2015gridgraph, cheng2015venus, chi2016nxgraph, ai2017squeezing, ma2017garaph, maass2017mosaic, wu2018redio, ai2018clip, jun2018grafboost, zhang2018wonderland}} \cite{kyrola2012graphchi, han2013turbograph, roy2013x, najeebullah2014bishard, lin2014mmap, zheng2015flashgraph, wang2021scaleg} focus on loading large graphs stored in binary formats. However, a majority of graph datasets exist in serialized human-readable data exchange formats.\ignore{To address this, our focus lies on efficiently loading graphs stored in plain text formats.}

To address these challenges, Gabert and Çatalyürek introduce PIGO \cite{gabert2021pigo}, a header-only, dependency-free C++11 parallel graph loader that supports loading graphs in memory as Edgelists or CSR. PIGO leverages memory mapping, a mechanism that maps a file or part of a file into the virtual memory space\ignore{so that files on the disk can be accessed as if they were in memory} \cite{lin2014mmap}, to optimize file reading. This eliminates the need for repeated system calls, resulting in reduced context-switch overhead and improved efficiency, particularly if the kernel\ignore{accurately} predicts the accessed pages ahead of time.

However, we have identified a few issues with PIGO. Firstly, when reading entries from the input file, PIGO divides the file length equally among threads, potentially leading to slower overall performance as faster threads wait for slower ones. Secondly, PIGO utilizes a two-pass approach for loading graphs into memory as Edgelists, which involves first counting newlines to determine the number of edges and associated offsets for each thread, and then parsing and populating the Edgelist. This method is inefficient compared to a single-pass approach. When converting the Edgelist to a CSR representation, PIGO globally computes vertex degrees using atomics, uses it compute the offsets array of the global CSR, and iterates through the Edgelist to atomically populate the targets array of the global CSR. This global computation of vertex degrees, and directly operating on the shared CSR can lead to high contention between threads, especially on graphs with high average degree. Further, PIGO populates the targets array of the CSR with static load balancing, potentially leading to load imbalances among threads. Finally, when reading Matrix Market (MTX) files, a format commonly used for storing sparse graphs/matrices, PIGO disregards specified attributes, resulting in lower reported runtimes for symmetric graphs (the authors of PIGO plan to address this\ignore{in the future}).

In this technical report, we propose GVEL\footnote{\url{https://github.com/puzzlef/graph-csr-openmp}}. Similar to PIGO, it employs memory mapping and parallelization to optimize graph loading. However, GVEL improves upon PIGO by efficiently processing the graph as per-thread Edgelists in a single pass through overallocation of memory via memory mapping. Note that this does not waste memory, as untouched pages are never mapped to DRAM. To convert the per-thead Edgelists to CSR, GVEL computes four independent sets of vertex degrees (which, when summed up for each vertex, represents the global degree of each vertex), and uses it to generate the global CSR representation in a novel staged manner. It does this by first obtaining $4$ independent sets of CSRs, and then combining them together, in parallel, to form a global CSR. This minimizes the contention between threads\ignore{, which we observe to be a significant bottleneck for converting an Edgelist to CSR}. These techniques allow GVEL to achieve a $2.6\times$ speedup over PIGO for loading graphs into memory as Edgelists, and a speedup of $1.8\times$ for loading graphs into memory as CSRs (i.e., reading the graph as per-thread Edgelists and then converting it to CSR). Our techniques may also be used to convert in-memory Edgelists (an update friendly data structure), to a CSR (a space efficient and locality efficient data structure).

\section{Related work}
\label{sec:related}
Graph processing frameworks are software systems designed to efficiently analyze and manipulate graph-structured data, facilitating tasks such as traversal, analytics, and algorithm implementation on various computational architectures. Ligra \cite{shun2013ligra} is a lightweight framework designed for shared-memory parallel/multicore machines, simplifying the development of graph traversal algorithms with two simple routines for edge and vertex mapping. The Galois system \cite{nguyen2013lightweight} advocates for a general-purpose infrastructure supporting fine-grain tasks, speculative execution, and application-specific task scheduling control. Gunrock \cite{wang2016gunrock} tackles irregular data access and control flow issues on GPUs by introducing a data-centric abstraction focused on vertex or edge frontier operations, and enables rapid development of new graph primitives with minimal GPU programming knowledge. Hornet \cite{busato2018hornet} provides a scalable GPU implementation without requiring data reallocation during evolution, targeting dynamic data problems. While frameworks such as these have demonstrated efficient computation on billion-scale graphs, many of them persist with sequential I/O.

Hornet \cite{busato2018hornet} reads a graph in the Matrix Market (MTX) format using the stream operator on \texttt{ifstream}, and then converts it to CSR using \texttt{COOtoCSR()} (in the \texttt{graph::GraphBase::read()} function). The \texttt{COOtoCSR()} function then calculates the out-degrees of each vertex and computes the prefix sum (using \texttt{std::partial\_sum()}) to determine the offsets of outgoing edges in the CSR representation, fills in the CSR arrays ($\_out\_edges$ and $\_out\_offsets$) based on the sorted or randomized COO edges, and if the graph is directed, it also calculates the in-degrees and fills in the corresponding arrays ($\_in\_edges$ and $\_in\_offsets$). In Gunrock \cite{wang2016gunrock}, the \texttt{io::matrix\_ma} \texttt{rket\_t::load()} function handles reading of graphs in the MTX format. It uses \texttt{fscanf()} to read the entries from the input file into three separate \texttt{vector}s, which are then passed to the \texttt{format::csr} \texttt{\_t::from\_coo()} function to generate a CSR. In GraphBLAST \cite{yang2022graphblast}, the \texttt{readMtx()} function is used to read graphs in the MTX format. It uses \texttt{fscanf()} to read tuples/entries, and pushes then to a \texttt{vector}. It then removes self-loops, and does a custom sort of the entries. The GAP Benchmark Suite \cite{beamer2015gap} uses \texttt{ifstream} with \texttt{getline()} and stream operator to read into COO. In each of these frameworks, all of the operations are performed sequentially.

Incorporating Memory-Mapped I/O holds promise for improving graph loading efficiency of frameworks. A number of research studies have worked on optimizing the performance of memory mapped IO. The approaches include using a lightweight userspace memory service \cite{li2019userland}, huge pages \cite{malliotakis2021hugemap}, and asynchronous techniques \cite{imamura2019poster}. A\ignore{large} number of works have optimized memory mapped IO for fast low-latency storage devices \cite{song2012low, song2016efficient, papagiannis2020optimizing, papagiannis2021memory, alverti2022daxvm, leis2023virtual}. Essen et al. \cite{van2015di} and Feng et al. \cite{feng2023tricache} focus on enabling programs to efficiently process out-of-core datasets through memory mapping.

In NetworKit \cite{staudt2016networkit}, the \texttt{MatrixMarketReader::read()} function handles reading of graphs in the MTX format. The function uses \texttt{istream} to read each entry using the stream operator, which is then pushed to a \texttt{vector}. This vector is then passed to \texttt{CSRMatrix}, which sequentially adds the entries in the vector to a CSR. Since the offsets array of the CSR must start with a zero, it puts a zero at the end and rotates the offsets array. We now discuss the \texttt{EdgeListReader::re} \texttt{ad()} function, which handles reading of graphs in COO/Edgelist format, and returns a \texttt{Graph}. It maps the file to memory with the \texttt{MemoryMappedFile} class, reads source/target vertex IDs with \texttt{strtol()}, edge weights with \texttt{strtod()}, conditionally adjusts the number of nodes at each step with \texttt{graph.addNodes()}, and conditionally adds an edge to the graph with \texttt{graph.addEdge()}. However, the function sequential, and adding each edge to the graph is expensive, i.e., $O(D)$ (where $D$ is the average degree). Further, the \texttt{MemoryMappedFile} class does not use \texttt{madvice()} to recommend a memory access pattern to the OS.\ignore{The call to \texttt{mmap()} uses \texttt{PROT\_READ} and \texttt{MAP\_PRIVATE}.}

Gabert and Çatalyürek \cite{gabert2021pigo} observe that even on high-performance shared-memory graph systems running billion-scale graphs, reading the graph from file systems takes multiple orders of magnitude longer than running the computational kernel. To address this issue, they propose PIGO, a high-performance parallel graph loader that brings I/O improvements to graph systems.

In the \texttt{COO::read\_mm\_()} function, PIGO handles the reading of graphs in MTX format. It employs the \texttt{O\_DIRECT} flag in Linux to open the file and maps the entire file to memory using \texttt{mmap()}, with \texttt{MAP\_SHARED} and \texttt{MAP\_NORESERVE} flags. It additionally utilizes \texttt{madvice()} with the \texttt{MADV\_WILLNEED} flag. PIGO however disregards MTX attributes. The authors of PIGO plan to fix this in the future. It then reads the first line of the file, containing information about rows, columns, and the number of edges. For integer reading, it navigates the file reader pointer to the next digit, reads an integer (using custom code), and then moves to the end-of-line. PIGO then proceeds to read the Edgelist with the \texttt{COO::read\_el\_()} function. Here, the split is split into equally sized parts for each thread to process. Each thread then adjusts the boundaries of its part, in order to eliminate overlapping entries. Now, in \texttt{COO::read\_el\_()}, a two pass approach is employed. In the first pass, each thread simply counts newlines in the part of file assigned to it and performs a prefix sum to determine write addresses for each thread into the Edgelist. In the second pass, PIGO iterates through the input file again, but this time, parsing and populating the source/target vertex IDs into the Edgelist. Memory allocation for \texttt{src}, \texttt{dst}, and \texttt{weights} is performed separately using \texttt{std::vector::resize()}.

After reading the input graph as an Edgelist, PIGO can convert it to the CSR format using the \texttt{convert\_coo\_()} function, or the \texttt{read\_graph\_()} function if the file type is specified as \texttt{GRAPH}. In \texttt{convert\_coo\_()}, PIGO allocates space for CSR and employs a multi-pass algorithm. Initially, it computes the each vertex's degree in a shared global array, and then transforms the degrees into the offsets array of the CSR using an exclusive sum algorithm. Then, treating the computed degrees as remaining vertices, Finally, PIGO copies over the edges into the targets array of the CSR. It does this with static loop scheduling, and atomic operations.\ignore{Temporary arrays are subsequently freed.}

\ignore{cuDF employs a GPU-accelerated parsing algorithm to efficiently read and interpret the CSV data within each chunk. This algorithm is optimized for GPU architecture and takes advantage of parallelism to process data quickly.}

\section{Preliminaries}
\label{sec:preliminaries}
\subsection{Graph Storage and In-Memory Formats}

Graphs can be stored in either text or binary formats. Text formats, such as Edgelists, provide a simple representation where each line denotes an edge between two vertices (optionally including edge weights and other attributes). While text formats are space-inefficient compared to binary formats, their readability and ease of sharing contribute to their widespread adoption in the public domain. Matrix Market (MTX) is a standardized text format with headers specifying the matrix's properties, making it suitable for various sparse matrix representations.

In-memory graph formats, on the other hand, optimize for efficient data access during computation. These include Edgelists, which mirror the structure of the Edgelist storage format. To improve data locality, the source vertex, target vertex, and edge weights may be stored in separate arrays. This format is easily convertible from Edgelist storage, and is suitable for graph algorithms requiring edge-oriented computation. Compressed Sparse Row (CSR) is another popular in-memory format that is optimal for vertex-oriented algorithms, such as traversal. It stores graph data in three arrays: one for offsets to the list of target vertices / edge weights of each vertex, one for target vertices (for each vertex), and one for edge weights (for each vertex).

\subsection{Memory-Mapped I/O}

Memory-mapped I/O (MMIO) is an access method that maps files or file-like resources to a memory region, providing applications with data access through memory semantics. \texttt{mmap()}, a key system call in memory-mapped I/O, is used to map files into memory, establishing a virtual memory mapping between the process's address space and the file or device. When a thread accesses a page that has not yet been loaded from the file, a page fault occurs, and the thread is put into sleep by the kernel until data from the file has been loaded into the page. The \texttt{madvise()} system call allows programmers to advise the kernel about their expected access patterns for the mapped region, optimizing performance. Subsequently, \texttt{munmap()} is employed to unmap the region, freeing up resources.

MMIO has reduced overhead and eliminates copies between kernel and user space \cite{malliotakis2021hugemap, papagiannis2020optimizing}. Hot data typically resides in main memory, leveraging the benefits of a large cache, while cold data are stored on devices such as HDDs and SSDs \cite{song2016efficient}. The increasing adoption of MMIO is driven by its superior performance, particularly in comparison to file semantics like read/write operations which introduces significant overhead for fast storage \cite{yoshimura2019evfs, enberg2022transcending}. 

\ignore{Memory mapping is a mechanism that maps a file or part of a file into the virtual memory space, so that files on the disk can be accessed as if they were in memory \cite{lin2014mmap}. There are many advantages of using memory mapping, especially when processing large files. - Reading from and writing to a memory-mapped file do not require the data to be copied to and from a user-space buffer while standard read/write do. - Aside from any potential page faults, reading from and writing to a memory-mapped file do not incur any overhead due to context switching. - When multiple processes map the same data into memory, they can access that data simultaneously. Read-only and shared writable mappings are shared in their entirety; private writable mappings may have their not-yet-COW (copy-on-write) pages shared \cite{lin2014mmap}.}

\subsection{Stream Operations in C++}

The Standard Template Library (STL) in C++ offers a suite of stream classes and methods. C++ stream classes include \texttt{ifstream} for input file streams and \texttt{ofstream} for output file streams. The \texttt{istream} and \texttt{ostream} interfaces provide a common set of methods for reading and writing data. The \texttt{>>} and \texttt{<<} operators are overloaded for various data types and provide read and writing of data from and to streams, while \texttt{getline()} method reads a line from the stream. C++ streams use an internal buffer to optimize I/O operations.

\subsection{Number Parsing Methods in C++}

C++ provides several methods for converting textual data to numerical representations. \texttt{sscanf()} is a powerful method for parsing formatted input strings. It operates similarly to `printf` and allows developers to define a format specifier that describes the expected structure of the input string. On the other hand, \texttt{strtoull()} and \texttt{strtod()} offer a simple and direct approach for converting strings to unsigned long long integers and double-precision floating-point numbers, respectively. Performance-wise, \texttt{sscanf()} might involve a bit more overhead due to the need to interpret format specifiers, while \texttt{strtoull} and \texttt{strtod} provide straightforward conversions without the need for format strings.

\section{Approach}
\label{sec:approach}
\subsection{Reading Edgelist from text file}

We attempt a number of approaches to read Edgelist from text file into in-memory Edgelist(s), given in Sections \ref{sec:el-fstream-plain}-\ref{sec:el-mmap-custom}.

\subsubsection{\texttt{ifstream} with \texttt{getline()} and \texttt{>>} operator (\textit{fstream-plain})}
\label{sec:el-fstream-plain}

In this approach, we utilize C++'s \texttt{ifstream} to open the file, and read the edges line by line with \texttt{getline()}. If the graph is unweighted, we read the source and target vertex ids as 64-bit unsigned integers, using \texttt{>>} operator, into pre-allocated source and target arrays based on information in the file header. If the graph is weighted, we also read the weights as 32-bit floating-point numbers into another pre-allocated array. This process is sequential, given that streams are inherently sequential.

\subsubsection{\texttt{ifstream} with \texttt{getline()} and \texttt{strto*()} (\textit{fstream-strto*})}
\label{sec:el-fstream-stro*}

\begin{figure}[hbtp]
  \centering
  \includegraphics[width=0.99\linewidth]{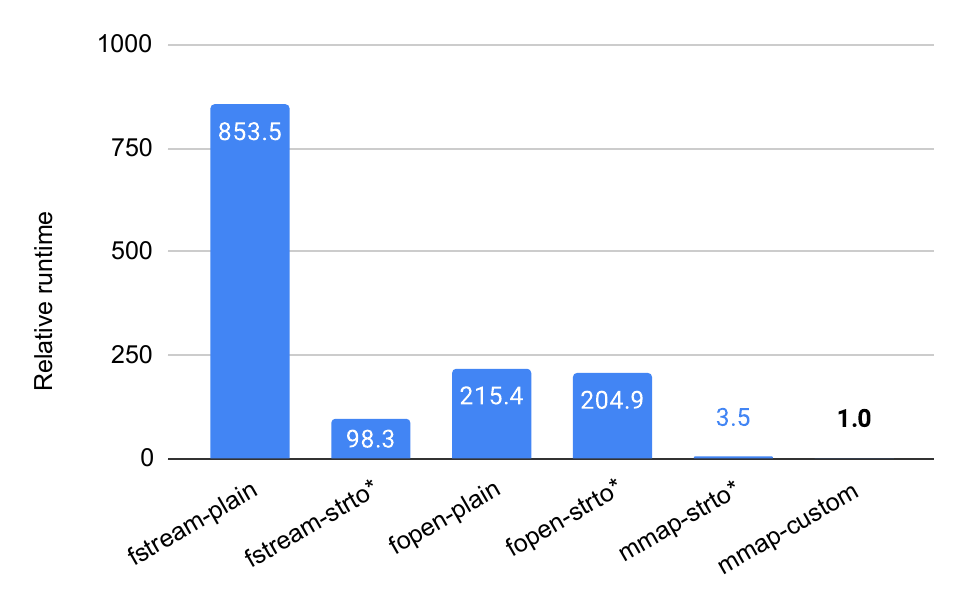} \\[-2ex]
  \caption{Relative runtime of reading per-thread Edgelists using C++'s input file stream (\textit{fstream-plain}), file stream to read lines and using  \texttt{strtoull} and \texttt{strtod} function for parsing numbers (\textit{fstream-strto*}), \texttt{fgets} to read lines and \texttt{sscanf} for parsing numbers (\textit{fopen-plain}), \texttt{fgets} with \texttt{strtoull} and \texttt{strtod} (\textit{fopen-strto*}), memory mapped file using \texttt{mmap} with \texttt{strtoull} and \texttt{strtod} (\textit{mmap-strto*}), and \texttt{mmap} with custom integer/floating-point parsing functions along with making vertex id 0-based (\textit{mmap-custom}).}
  \label{fig:optimize-el}
\end{figure}

In this approach, we again use \texttt{ifstream} but employ string-to-number conversion methods \texttt{strtoull()} and \texttt{strtod()} for parallel number parsing. We sequentially read a block of $L$ lines from the file, using \texttt{getline()}, and then parse each line in parallel using multiple threads. We observe that using OpenMP's dynamic scheduling, with a chunk size of $1024$, and reading a block of $L=128K$ lines to be processed in parallel offers the best performance. Parsed edges (source, target vertex ids, and edge weights) are stored separately in per-thread edge lists to avoid contention issues within a shared data structure. With 64 threads, this approach demonstrates a speedup of $8.7\times$ compared to \textit{fstream-plain}, as shown in Figure \ref{fig:optimize-el}.

\subsubsection{\texttt{fopen()} with \texttt{fgets()} and \texttt{sscanf()} (\textit{fopen-plain})}
\label{sec:el-fopen-plain}

This approach is similar to the one mentioned above (\textit{fscanf-strto*}), but we use \texttt{fgets()} on a file handle to read lines instead of \texttt{getline()}, and employ \texttt{sscanf()} to parse the edges. With 64 threads, it provided a speedup of $4.0\times$ compared to \textit{fstream-plain}.

\subsubsection{\texttt{fopen()} with \texttt{fgets()} and \texttt{strto*()} (\textit{fopen-strto*})}
\label{sec:el-fopen-strto*}

Similar to the previous approach (\textit{fopen-plain}), this one uses \texttt{fgets()} to read lines from the text file, but replaces \texttt{sscanf()} with \texttt{strtoull()} and \texttt{strtod()}. This proves faster due to the absence of a format string. At 64 threads, its speedup is $1.1\times$ over \textit{fopen-plain}.

\subsubsection{\texttt{mmap()} with \texttt{strto*()} (\textit{mmap-strto*})}
\label{sec:el-mmap-strto*}

\begin{figure}[hbtp]
  \centering
  \includegraphics[width=0.99\linewidth]{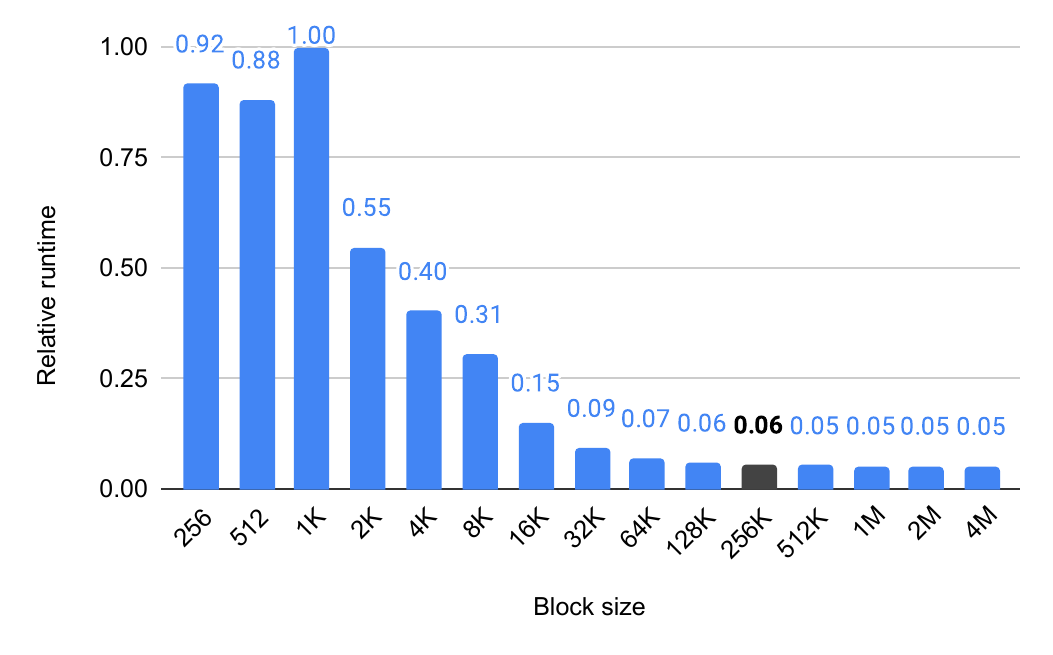} \\[-2ex]
  \caption{Relative runtime of reading per-thread Edgelists with varying block sizes (each thread is dynamically assigned a block of characters to process), from $256$ bytes to $4M$ bytes, using \texttt{mmap} with custom integer/float parsing functions along with making vertex id 0-based (\textit{mmap-custom}).}
  \label{fig:adjust-blocksize}
\end{figure}

In this approach, we map the file to memory with \texttt{mmap()}, use \texttt{madvice(MADV\_WILLNEED)} indicating the kernel to read-ahead pages, and process the edges in parallel by partitioning the file into blocks of $C$ characters. Each block is dynamically assigned (using OpenMP's dynamic schedule) to a free thread. If the assigned block contains partial lines at either end, the thread repositions it, by shifting to the right to eliminate partial lines. This involves skipping the partial line at the beginning and including the partial line from the end. We experiment with block sizes $\beta$ ranging from $256$ to $4M$ characters. Figure \ref{fig:adjust-blocksize} illustrates the average relative runtime for each block size across all graphs in Table \ref{tab:dataset}. As the figure shows, larger block sizes generally offer improved performance, but returns diminish beyond a certain block size. Further, choosing an extremely large block size can minimize the amount of parallelism for mid-sized graphs. Accordingly, we select a block size of $\beta=256K$ characters. To parse the source/target vertex ids and edge weights, we use \texttt{strtoull()} and \texttt{strtod()}. Each thread stores the parsed edges in per-thread Edgelists. As shown in Figure \ref{fig:optimize-el}, with 64 threads, this approach provides a speedup of $58.5\times$ over \textit{fopen-strto*}.

\subsubsection{\texttt{mmap()} with custom number parsers (\textit{mmap-custom})}
\label{sec:el-mmap-custom}

This is similar to the approach mentioned above (\textit{mmap-strto*}), but we use our own functions for parsing whole numbers and floating-point numbers. In addition, as vertex ids start with $1$, we decrement $1$ from the vertex ids after parsing it and before appending them to per-thread Edgelists. Surprisingly, this leads to $40-50\%$ drop in performance. Converting the $weighted$ flag (see Algorithm \ref{alg:el}) to a template parameter solves this issue. This indicates that the issue was related to the loop code not being able to fit in the code cache of the processor and using a template allowed it to fit in the cache. Accordingly, we also recommend using $symmetric$ flag as a template parameter instead. With 64 threads, it provides a speedup of $3.5\times$ over \textit{mmap-strto*}. We also attempted to use custom SIMD instructions to parse numbers, along with \texttt{vzeroupper} instruction to clear SSE/AVX registers, but it did not provide additional performance\ignore{improvement}.

Among the approaches given in Sections \ref{sec:el-fstream-plain}-\ref{sec:el-mmap-custom}, we find using \texttt{mmap()} with custom number parsers (\textit{mmap-custom}) to be the best approach. The pseudocode for \textit{mmap-custom} is given Algorithm \ref{alg:el}. It loads per-thread Edgelists from a file with the best performance, and is enacpsulated in the \texttt{readEdgelist()} function (lines \ref{alg:el--read-edgelist-begin}-\ref{alg:el--read-edgelist-end}). First, the counts of edges read per thread ($counts$) are initialized, and the components of per-thread Edgelists are obtained ($edges$) in lines \ref{alg:el--initialize-begin}-\ref{alg:el--initialize-end}. This is followed by a loop (lines \ref{alg:el--blocks-begin}-\ref{alg:el--blocks-end}), where each iteration processes a block of characters (of size $\beta = 256K$) in the text file in parallel across different threads. The text file is assumed to have been memory mapped as $data$. Inside the loop, $j$ keeps track of the number of edges processed by the current thread. In line \ref{alg:el--get-block}, the \texttt{getBlock()} function is called to retrieve the begin and end of current block of text ($[b, B]$), which is processed in the main loop (lines \ref{alg:el--block-begin}-\ref{alg:el--block-end}). The main loop parses edges (lines \ref{alg:el--parse-edge-begin}-\ref{alg:el--parse-edge-end}), adjusts vertex indices to be zero-based (line \ref{alg:el--base1}), and adds them to the Edgelist of the current thread while updating per-partition vertex degrees with atomic operations (lines \ref{alg:el--add-edge-begin}-\ref{alg:el--add-edge-end}). Reverse edges are added for symmetric graphs (lines \ref{alg:el--reverse-edge-begin}-\ref{alg:el--reverse-edge-end}). Finally, counts of processed edges per thread are recorded (line \ref{alg:el--update-counts}) and returned (line \ref{alg:el--return-counts}).

The \texttt{getBlock()} function (lines \ref{alg:el--get-block-begin}-\ref{alg:el--get-block-end}) retrieves a block of characters to process from the memory-mapped file, starting from index $i$. It ensures that the block starts and ends on newline characters for proper parsing. The block size is determined by the parameter $\beta$, which is set to $256K$.

\begin{algorithm}[hbtp]
\caption{Reading Edgelist from file.}
\label{alg:el}
\begin{algorithmic}[1]
\Require{$pdegrees$: Per partition vertex degrees (output)}
\Require{$edges$: Per thread sources, targets, and weights of edges (output)}
\Require{$data$: Memory mapped file data}
\Ensure{$counts$: Number of edges read per thread (output)}
\Ensure{$symmetric$: Is graph symmetric?}
\Ensure{$weighted$: Is graph weighted?}
\Ensure{$\beta$: Size of each block that is processed per thread}
\Ensure{$\rho$: Number of partitions for counting vertex degrees}
\Ensure{$t$: Current thread}

\Statex

\Function{readEdgelist}{$pdegrees, edges, data$} \label{alg:el--read-edgelist-begin}
  \State $counts \gets \{0\}$ \label{alg:el--initialize-begin}
  \State $[sources, targets, weights] \gets edges$ \label{alg:el--initialize-end}
  \State $\rhd$ Load edges from text file in blocks of size $\beta$
  \ForAll{$i \in [0, \beta, 2\beta, ... |data|]$ \textbf{in parallel}} \label{alg:el--blocks-begin}
    \State $j \gets counts[t]$
    \State $[b, B] \gets getBlock(data, i)$ \label{alg:el--get-block}
    \While{$true$} \label{alg:el--block-begin}
      \State $\rhd$ Read an edge from the block
      \State $u \gets v \gets 0$ \textbf{;} $w \gets 1$ \label{alg:el--parse-edge-begin}
      \State $b \gets findNextDigit(b, B)$
      \If{$b = B$} \textbf{break}
      \EndIf
      \State $b \gets parseWholeNumber(u, b, B)$
      \State $b \gets findNextDigit(b, B)$
      \State $b \gets parseWholeNumber(v, b, B)$
      \If{$weighted$}
        \State $b \gets findNextDigit(b, B)$
        \State $b \gets parseFloat(w, b, B)$
      \EndIf \label{alg:el--parse-edge-end}
      \State $\rhd$ Make it zero-based
      \State $u \gets u - 1$ \textbf{;} $v \gets v - 1$ \label{alg:el--base1}
      \State $\rhd$ Add the parsed edge to edgelist
      \State $sources[t][j] \gets u$ \label{alg:el--add-edge-begin}
      \State $targets[t][j] \gets v$
      \If{$weighted$} $weights[t][j] \gets w$
      \EndIf
      \State $atomicAdd(pdegrees[t \bmod \rho][u], 1)$ \label{alg:el--update-degrees}
      \State $j \gets j + 1$ \label{alg:el--add-edge-end}
      \State $\rhd$ If graph is symmetric, add the reverse edge
      \If{$symmetric$} \label{alg:el--reverse-edge-begin}
        \State $sources[t][j] \gets v$
        \State $targets[t][j] \gets u$
        \If{$weighted$} $weights[t][j] \gets w$
        \EndIf
        \State $atomicAdd(pdegrees[t \bmod \rho][v], 1)$
        \State $j \gets j + 1$
      \EndIf \label{alg:el--reverse-edge-end}
    \EndWhile \label{alg:el--block-end}
    \State $counts[t] \gets j$ \label{alg:el--update-counts}
  \EndFor \label{alg:el--blocks-end}
  \Return{$counts$} \label{alg:el--return-counts}
\EndFunction \label{alg:el--read-edgelist-end}

\Statex

\Function{getBlock}{$data, i$} \label{alg:el--get-block-begin}
  \State $[d, D] \gets data$
  \State $b \gets d+i$ \textbf{;} $B \gets min(b+\beta, D)$
  \If{$b \neq d$ \textbf{and not} $isNewline(b-1)$}
    \State $b \gets findNextLine(b, D)$
  \EndIf
  \If{$B \neq d$ \textbf{and not} $isNewline(B-1)$}
    \State $B \gets findNextLine(B, D)$
  \EndIf
  \Return{$[b, B]$}
\EndFunction \label{alg:el--get-block-end}
\end{algorithmic}
\end{algorithm}

\subsection{Converting Edgelist to CSR}

Now that we have obtained per-thread Edgelists, we must now convert the Edgelists to CSR. We attempt a few of approaches for this in steps, given in Sections \ref{sec:csr-degree-global}-\ref{sec:csr-csr-partition4}.

\subsubsection{Obtain global vertex degrees along with reading Edgelist (\textit{degree-global})}
\label{sec:csr-degree-global}

To convert the Edgelists to CSR, we first need to know the degree of each vertex. In this approach, we use a simple solution for this, i.e.,  we update the degree of each vertex in a single shared array using atomic operations, while reading Edgelists on each thread. The relative runtime of this approach with respect to simply reading per-thread Edgelists is $1.9\times$, as shown in Figure \ref{fig:optimize-csr}. We however observe that this results in high contention and negatively impacts performance.

\subsubsection{Obtain per-thread vertex degrees along with reading Edgelist (\textit{degree-thread})}
\label{sec:csr-degree-thread}

In this approach, we compute per-thread vertex degrees instead of global degrees. While this improves performance by $24\%$ compared to obtaining global vertex degrees, it requires significant additional space, and needs to be combined later to obtain global degrees. We observe that computing degrees in partitions of $4$ (using $\bmod 4$) and then combining them into a global degrees array gives us the best performance.

\subsubsection{Obtain CSR from global vertex degrees (\textit{csr-global})}
\label{sec:csr-csr-global}

\begin{figure}[hbtp]
  \centering
  \includegraphics[width=0.99\linewidth]{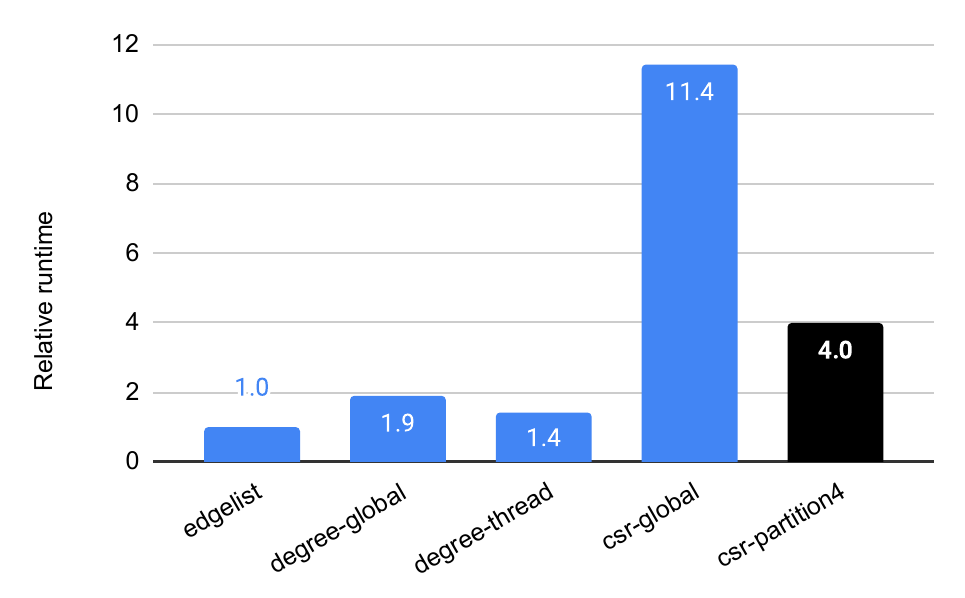} \\[-2ex]
  \caption{Relative runtime of reading per-thread Edgelists from file (\textit{edgelist}), reading per-thread Edgelists + global degrees (\textit{degree-global}), reading per-thread Edgelists + per-thread degrees (\textit{degree-thread}), reading per-thread Edgelists + global degrees + converting to global CSR (\textit{csr-global}), and reading per-thread Edgelists + 4-partition degrees + converting to 4-partition CSR + converting to global CSR (\textit{csr-partition-4}).}
  \label{fig:optimize-csr}
\end{figure}

Now that we have the degrees, we need to combine the per-thread Edgelists into a CSR\ignore{data structure}. In this approach, we first obtain global vertex degrees along with reading per-thread Edgelists (as with \textit{degree-global}), and then convert the per-thread Edgelists to a global CSR in parallel using atomic operations. As shown in Figure \ref{fig:optimize-csr}, this\ignore{approach} has poor performance (it is $11.4\times$ slower than simply reading Edgelists).

\begin{figure}[hbtp]
  \centering
  \subfigure{
    \label{fig:adjust-csrpartitions--full}
    \includegraphics[width=0.98\linewidth]{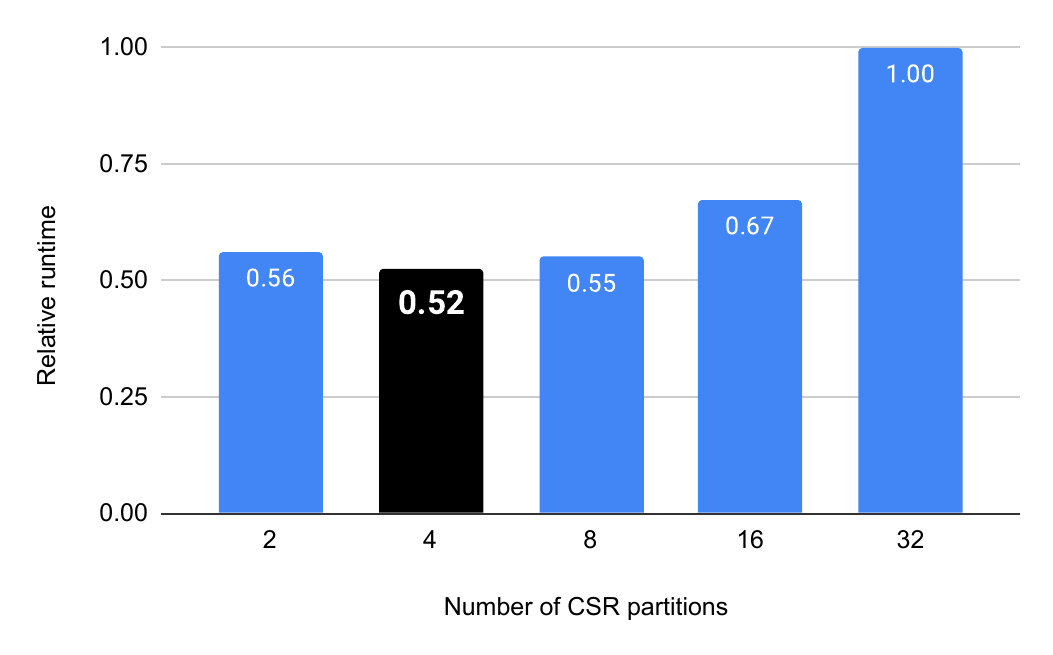}
  } \\[-2ex]
  \caption{Relative runtime of reading per-thread Edgelists, obtaining $k$-partition degrees, converting to $p$-partition CSR, and combining to global CSR, with the number of partitions $k$ varying from $2$ to $32$. Here, $k > 1$ implies a multi-stage approach for generating CSR, which minimizes contention\ignore{between threads}.}
  \label{fig:adjust-csrpartitions}
\end{figure}

\subsubsection{Obtain CSR from 4-partitioned vertex degrees (\textit{csr-partition4})}
\label{sec:csr-csr-partition4}

Finally, we explore computing CSR in $k$ separate partitions and later combining them into a single global CSR. The vertex degrees are also obtained in $k$ partitions while reading Edgelist, which is then used for generating partitioned CSR. We adjust the value of $k$ from $2$ to $32$.\ignore{Figure \ref{fig:adjust-csrpartitions--convert} shows the relative runtime of converting in-memory Edgelists to CSR, with varying values of $k$, while} Figure \ref{fig:adjust-csrpartitions} shows the relative runtime of the overall process, i.e., reading graphs as CSRs. As the figure indicates, using a multi-staged approach for generating CSR, where the first stage generates CSR in $4$ separate partitions, while the second stage combines them into a global CSR, has the best performance --- it offers $2.9\times$ speedup over \textit{csr-global}, as shown in Figure \ref{fig:optimize-csr}.

Among these, we find that obtaining CSR from vertex degrees, where the vertex degrees are computed in four separate partitions using $\bmod$ operator with the current thread id, performs the best. The pseudocode for \textit{csr-partition4} is given in Algorithm \ref{alg:csr}. It transforms an Edgelist representation of a graph into CSR format. First, the components of global CSR ($csr$), per-partition CSRs ($pcsr$), and per-thread Edgelists ($edges$) are obtained in lines \ref{alg:csr--initialize-begin}-\ref{alg:csr--initialize-end}. The algorithm then computes offsets for each partition in parallel (lines \ref{alg:csr--poffsets-begin}-\ref{alg:csr--poffsets-end}) using exclusive scan operations, and updates the per-partition CSRs concurrently (lines \ref{alg:csr--pcsr-begin}-\ref{alg:csr--pcsr-end}). Atomic operations ensure thread safety when updating the matrices. In lines \ref{alg:csr--poffsets-fix-begin}-\ref{alg:csr--poffsets-fix-end}, per-partition offsets are fixed as they were updated during the edge insertion process above. The algorithm then combines per-partition degrees (lines \ref{alg:csr--poffsets-combine-begin}-\ref{alg:csr--poffsets-combine-end}), computes global offsets (line \ref{alg:csr--offsets-compute}), and merges the per-partition CSRs into the global CSR (lines \ref{alg:csr--pcsr-combine-begin}-\ref{alg:csr--pcsr-combine-end}).

\begin{algorithm}[hbtp]
\caption{Convert per-thread Edgelists to CSR.}
\label{alg:csr}
\begin{algorithmic}[1]
\Require{$csr$: Global CSR (output)}
\Require{$pcsr$: Per partition CSR (scratch)}
\Require{$pdegrees$: Per partition vertex degrees (scratch)}
\Require{$edges$: Per thread sources, targets, and weights of edges}
\Require{$counts$: Number of edges read per thread}
\Ensure{$symmetric$: Is graph symmetric?}
\Ensure{$weighted$: Is graph weighted?}
\Ensure{$\rho$: Number of partitions for counting vertex degrees}
\Ensure{$t$: Current thread}

\Statex

\Function{convertToCsr}{$csr, pcsr, pdegrees, edges, counts$}
  \State $[offsets, edgeKeys, edgeValues] \gets csr$ \label{alg:csr--initialize-begin}
  \State $[poffsets, pedgeKeys, pedgeValues] \gets pcsr$
  \State $[sources, targets, weights] \gets edges$ \label{alg:csr--initialize-end}
  \State $\rhd$ Compute offsets
  \ForAll{$p \in [0, \rho)$} \label{alg:csr--poffsets-begin}
    \State $exclusiveScan(poffsets[p], pdegrees[p], |V|+1)$
  \EndFor \label{alg:csr--poffsets-end}
  \State $\rhd$ Populate per-partition CSR
  \ForAll{\textbf{threads in parallel}} \label{alg:csr--pcsr-begin}
    \ForAll{$i \in [0, counts[t])$}
      \State $u \gets sources[t][i]$
      \State $v \gets targets[t][i]$
      \State $j \gets atomicAdd(poffsets[t \bmod \rho][u], 1)$
      \State $pedgeKeys[t \bmod \rho][j] \gets v$
      \If{$weighted$}
        \State $pedgeValues[t \bmod \rho][j] \gets weights[t][i]$
      \EndIf
    \EndFor
  \EndFor \label{alg:csr--pcsr-end}
  \State $\rhd$ Fix per-partition offsets
  \ForAll{\textbf{threads in parallel}} \label{alg:csr--poffsets-fix-begin}
    \If{$t < \rho$}
      \State $memcpy(poffsets[t]+1, poffsets[t], |V|)$
      \State $poffsets[t][0] \gets 0$
    \EndIf
  \EndFor \label{alg:csr--poffsets-fix-end}
  \State $\rhd$ Combine per-partition degrees
  \ForAll{$u \in [0, |V|)$ \textbf{in parallel}} \label{alg:csr--poffsets-combine-begin}
    \ForAll{$p \in [1, \rho)$}
      \State $pdegrees[0][u] +\gets pdegrees[p][u]$
    \EndFor
  \EndFor \label{alg:csr--poffsets-combine-end}
  \State $\rhd$ Compute global offsets
  \State $exclusiveScan(offsets, pdegrees[0], |V|+1)$ \label{alg:csr--offsets-compute}
  \State $\rhd$ Combine per-partition CSR into one CSR
  \ForAll{$u \in [0, |V|)$ \textbf{in parallel}} \label{alg:csr--pcsr-combine-begin}
    \State $j \gets offsets[u]$
    \ForAll{$p \in [0, \rho)$}
      \State $i \gets poffsets[t][u]$
      \State $I \gets poffsets[t][u+1]$
      \ForAll{$i \in [i, I)$}
        \State $edgeKeys[j] \gets pedgeKeys[t][i]$
        \If{$weighted$}
          \State $edgeValues[j] \gets pedgeValues[t][i]$
        \EndIf
        \State $j \gets j + 1$
      \EndFor
    \EndFor
  \EndFor \label{alg:csr--pcsr-combine-end}
\EndFunction
\end{algorithmic}
\end{algorithm}

\section{Evaluation}
\label{sec:evaluation}
\subsection{Experimental Setup}
\label{sec:setup}

We use a server that has two $16$-core x86-based Intel Xeon Gold 6226R processors running at $2.90$ GHz. Each core has an L1 cache of $1$ MB, an L2 cache of $16$ MB, and a shared L3 cache of $22$ MB. $12$ Seagate Exos 10e2400 $2.4$ TB SAS-12Gbps 2.5inch HDDs (DL2400MM0159), each with a rotational speed of $10k$ rpm, a sustained transfer rate of $241$ (outer diameter) to $117$ Mbps (inner diameter), and an average latency of $2.9$ ms, are installed on the server --- connected with a MegaRAID SAS-3 3008 SAS+SATA Controller. The machine has $512$ GB of system memory, and runs on CentOS Stream 8. We use GCC 8.5 and OpenMP 4.5. Table \ref{tab:dataset} shows the graphs we use in our experiments. All of them are obtained from the SuiteSparse Matrix Collection \cite{kolodziej2019suitesparse}.

\begin{table}[!ht]
  \centering
  \caption{In our experiments, we use a list of 13 graphs. The table lists the total number of vertices ($|V|$), total number of edges ($|E|$), and the file size ($F_{size}$) for each graph.}
  \label{tab:dataset}
  \begin{tabular}{|c||c|c|c|}
    \toprule
    \textbf{Graph} &
    \textbf{\textbf{$|V|$}} &
    \textbf{\textbf{$|E|$}} &
    \textbf{\textbf{$F_{size}$}} \\
    \midrule
    \multicolumn{4}{|c|}{\textbf{Web Graphs (LAW)}} \\ \hline
    indochina-2004 & 7.41M & 194M & 2.9 GB \\ \hline
    uk-2002 & 18.5M & 298M & 4.7 GB \\ \hline
    arabic-2005 & 22.7M & 640M & 11 GB \\ \hline
    uk-2005 & 39.5M & 936M & 16 GB \\ \hline
    webbase-2001 & 118M & 1.02B & 18 GB \\ \hline
    it-2004 & 41.3M & 1.15B & 19 GB \\ \hline
    sk-2005 & 50.6M & 1.95B & 33 GB \\ \hline
    \multicolumn{4}{|c|}{\textbf{Social Networks (SNAP)}} \\ \hline
    com-LiveJournal & 4.00M & 34.7M & 480 MB \\ \hline
    com-Orkut & 3.07M & 117M & 1.7 GB \\ \hline
    \multicolumn{4}{|c|}{\textbf{Road Networks (DIMACS10)}} \\ \hline
    asia\_osm & 12.0M & 12.7M & 200 MB \\ \hline
    europe\_osm & 50.9M & 54.1M & 910 MB \\ \hline
    \multicolumn{4}{|c|}{\textbf{Protein k-mer Graphs (GenBank)}} \\ \hline
    kmer\_A2a & 171M & 180M & 3.2 GB \\ \hline
    kmer\_V1r & 214M & 233M & 4.2 GB \\ \hline
  \bottomrule
  \end{tabular}
  \end{table}

\subsection{Performance Comparison for Reading CSR}

\begin{figure*}[hbtp]
  \centering
  \includegraphics[width=0.68\linewidth]{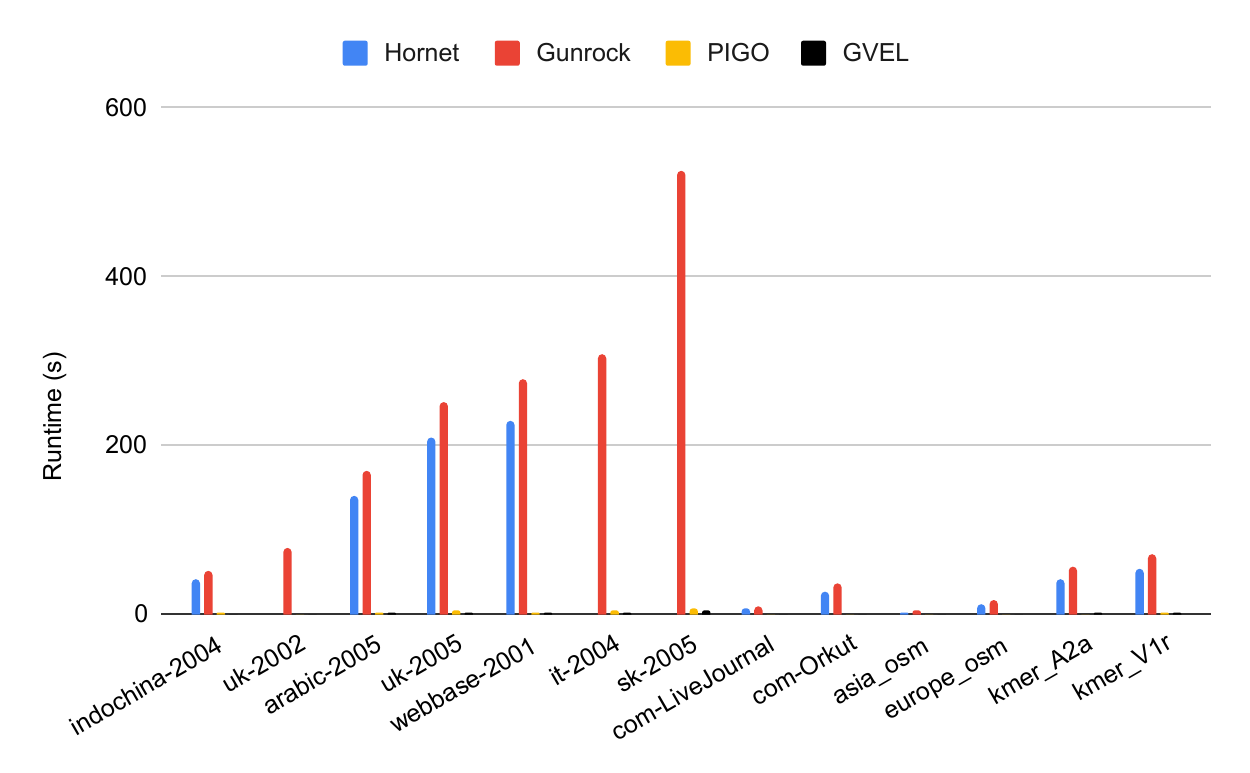} \\[-2ex]
  \caption{Time taken by Hornet, Gunrock, PIGO, and GVEL for reading Edgelist and converting it to CSR on 13 different graphs. PIGO and GVEL are not visible on this scale - they are significantly faster than Hornet and Gunrock. The graph loading time for Hornet is not shown for \textit{uk-2002}, \textit{it-2004}, and \textit{sk-2005} graphs as it crashed while loading.}
  \label{fig:compare-large}
\end{figure*}

\begin{figure*}[hbtp]
  \centering
  \includegraphics[width=0.68\linewidth]{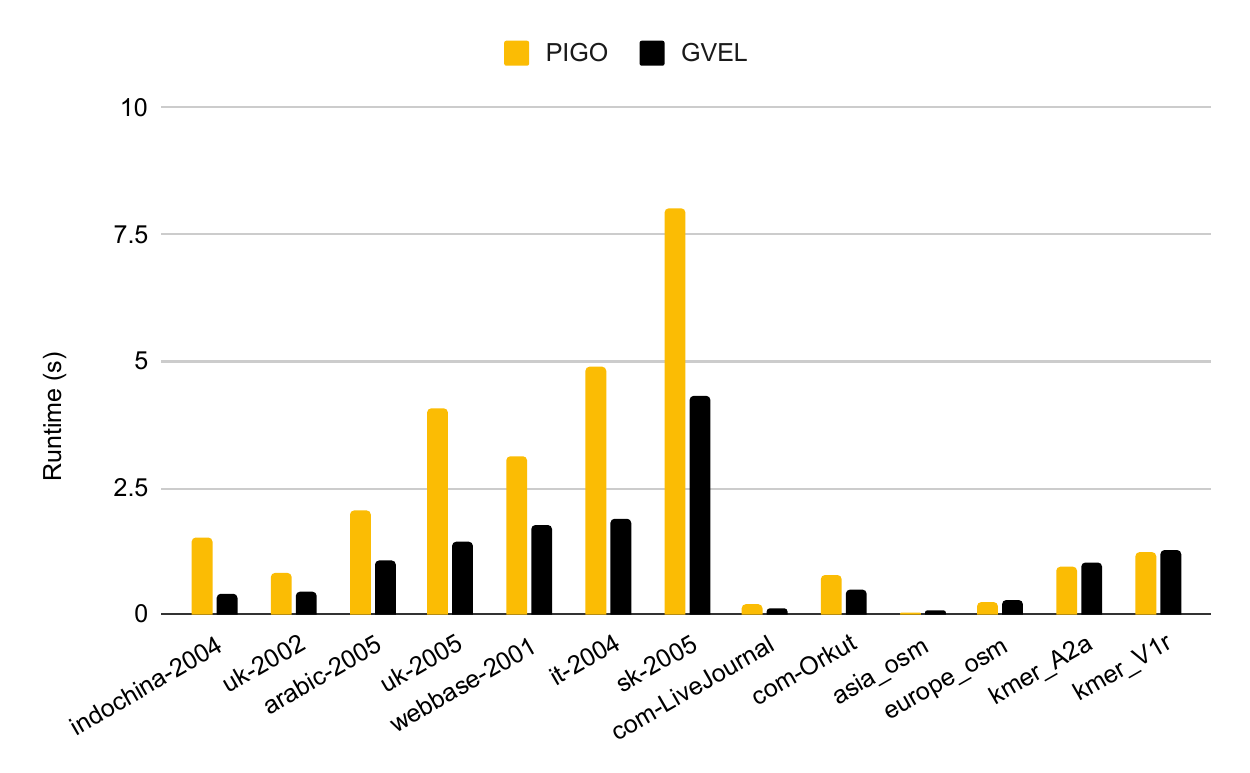} \\[-2ex]
  \caption{Time taken by PIGO and GVEL for reading Edgelist and converting it to CSR on 13 different graphs.}
  \label{fig:compare-small}
\end{figure*}

We now compare the performance of GVEL for reading CSR, i.e. reading Edgelist and converting it to CSR, with Hornet, Gunrock, and PIGO. For Hornet and Gunrock, we modify an example and/or source code of the framework to measure time taken to load the graph. For PIGO we include the released header-only file and write a program for measure CSR loading time. For each graph, we measure the CSR reading with each framework five times, for averaging. We also attempt to compare with GraphOne, but it appears to crash for all graphs except \textit{indochina-2004}. Figure \ref{fig:compare-large} shows the runtimes of Hornet, Gunrock, PIGO, and GVEL for reading CSR on each graph in the dataset. GVEL is on average $78\times$, $112\times$, and $1.8\times$ faster than Hornet, Gunrock, and PIGO respectively. The CSR reading time for Hornet is not shown for the graphs \textit{uk-2002}, \textit{it-2004}, and \textit{sk-2005} as it crashes while loading. Note that the runtimes of PIGO and GVEL are not visible on the scale of Figure \ref{fig:compare-large} as they are significantly faster than Hornet and Gunrock. Figure \ref{fig:compare-small} accordingly only shows the runtimes of PIGO and GVEL. Here, we observe that GVEL excels on web graphs, which are characterized by power law distributions and high average degrees. This advantage stems from our adoption of a multi-stage approach in constructing a CSR from the Edgelists, effectively minimizing contention among threads, particularly evident in high-degree graphs. Conversely, on graphs with low average degrees such as road networks and protein k-mer graphs, thread contention is minimal due to the low likelihood of simultaneous edge addition to the same vertex in the CSR targets array. Consequently, GVEL does not perform better than PIGO on such graphs. Further, PIGO only loads the upper-triangular part of undirected graphs (social networks, road networks, and protein k-mer graphs), i.e., it only loads half the number of edges in the graphs. This results in lower runtime being reported for PIGO.

\subsection{Performance Comparison for Reading Edgelist}

\begin{figure*}[hbtp]
  \centering
  \includegraphics[width=0.68\linewidth]{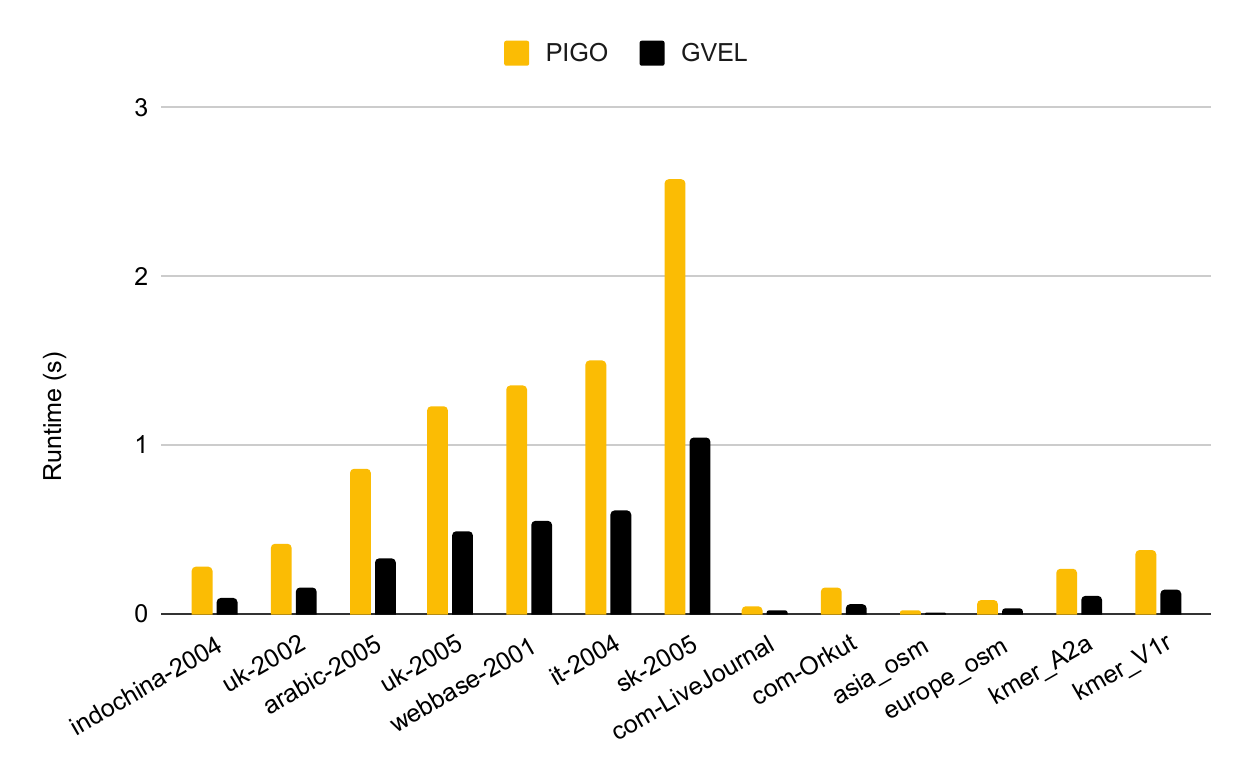} \\[-2ex]
  \caption{Time taken by PIGO and GVEL for reading Edgelist on 13 different graphs.}
  \label{fig:compare-el}
\end{figure*}

\begin{figure*}[hbtp]
  \centering
  \includegraphics[width=0.68\linewidth]{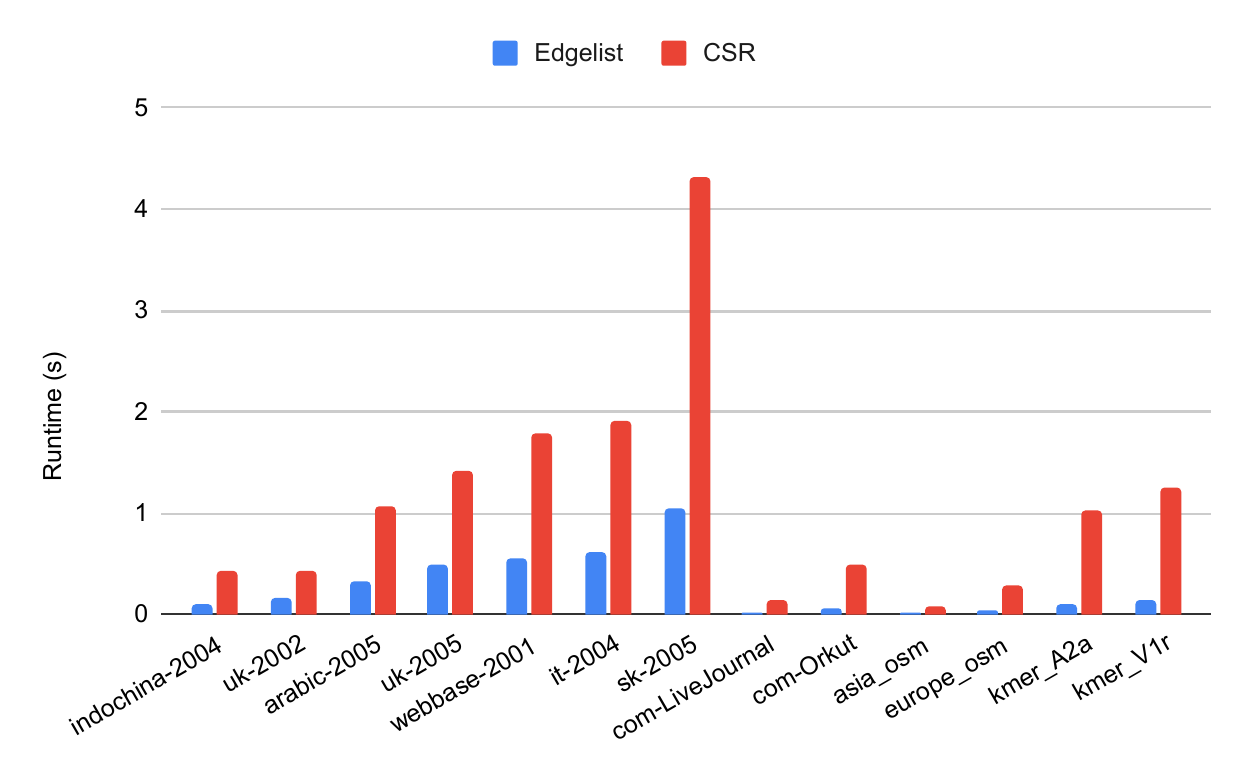} \\[-2ex]
  \caption{Time taken by GVEL for edge-list and CSR loading on 13 different graphs.}
  \label{fig:runtime}
\end{figure*}

In Figure \ref{fig:compare-el}, we compare the performance of GVEL for reading Edgelist with PIGO. As before, we measure the Edgelist reading time with each framework five times, for averaging. GVEL is on average $2.6\times$ faster than PIGO - with little skew, i.e., GVEL is more or less equivalently faster than PIGO on all graphs. This is more probably than not, due to the reading of Edgelist being \textit{pleasingly parallel}. On the \textit{sk-2005} graph, GVEL achieves a read performance of $1.9$ billion edges/s. Figure \ref{fig:runtime} illustrates the time GVEL requires to read the edge-list and convert it to CSR for each graph in the dataset, presented separately.

\subsection{Strong Scaling of GVEL}

\begin{figure*}[hbtp]
  \centering
  \includegraphics[width=0.68\linewidth]{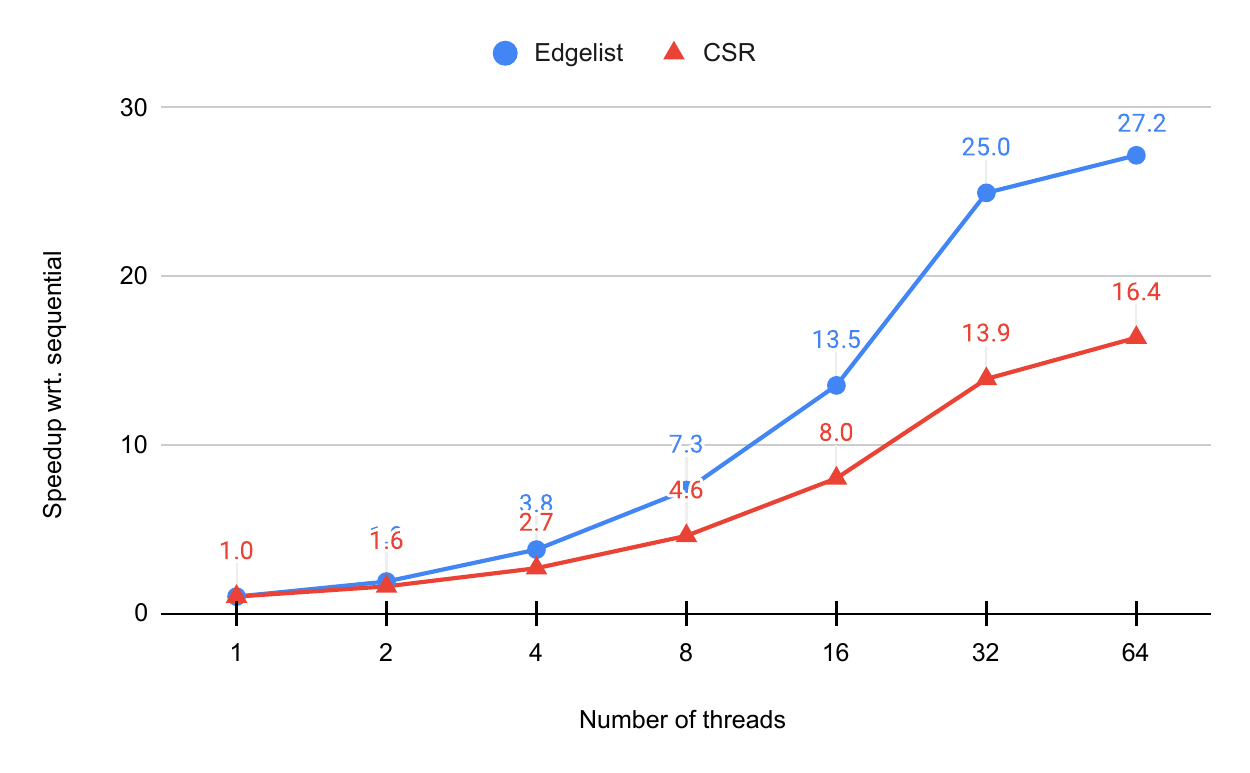} \\[-2ex]
  \caption{Speedup of GVEL's Edgelist and CSR loading with increasing number of threads.}
  \label{fig:strong-scaling}
\end{figure*}

Finally, we measure the strong scaling performance of GVEL. To this end, we adjust the number of thread from $1$ to $64$ in multiples of $2$ for each input graph, and measure the time taken for reading Edgelist and reading CSR (reading Edgelist + converting to CSR). As for each experiment above, for each graph, and for each thread count, we perform graph loading five times for averaging. The results are shown in Figure \ref{fig:strong-scaling}. With 32 threads, reading Edgelist with GVEL obtains a $25.0\times$ speedup compared to running with a single thread, i.e., the performance of Edgelist reading increases by $1.9\times$ for every doubling of threads. Further, with 32 threads, reading CSR with GVEL obtains a speedup of $13.9\times$ with respect to a single threaded execution, i.e., the CSR reading performance increases by $1.7\times$ for every doubling of threads. This drop in increase of speedup is due that fact that CSR is a shared data structure, leading to high contention when multiple threads concurrently attempt to add an edge to the same vertex, especially for vertices with high degrees. Further, the process of adding edges to vertices with small degrees is hampered by false sharing of cache lines, which contribute to decreased performance. At 64 threads, both reading Edgelist and reading CSR are impacted by NUMA effects, and offer speedups of only $27.2\times$ and $16.4\times$ respectively.

\section{Conclusion}
\label{sec:conclusion}
In conclusion, this study introduces GVEL, an optimized approach for efficient Edgelist reading and Compressed Sparse Row (CSR) conversion. Utilizing memory-mapped IO, a technique known for its efficiency in handling large datasets, GVEL outperforms established frameworks such as Hornet, Gunrock, and PIGO by $78\times$, $112\times$, and $1.8\times$ respectively, in CSR reading. For Edgelist reading, GVEL demonstrates a $2.6\times$ speedup over PIGO, and achieves a Edgelist read rate of $1.9$ billion edges/s. Our techniques may also be useful for converting in-memory Edgelists to CSRs on the fly on parallel devices. This is useful since CSR is an space and locality efficient data structure, while Edgelists are easy to update. Further, we observe that our novel multi-stage CSR construction, as employed in GVEL, is particularly effective on high-degree graphs, due to its ability of minimizing thread contention. However, on graphs with lower average vertex degrees, using the standard single-stage method of CSR construction, such as that used by PIGO, is the suitable approach.

\ignore{Looking ahead, future work could focus on extending GVEL to accommodate dynamic graphs which evolve over time, instead of CSR which is not amenable to dynamic updates. This would contribute to enhancing the adaptability and versatility of GVEL in dynamic graph processing scenarios.}

\begin{acks}
I would like to thank Prof. Kishore Kothapalli and Balavarun Pedapudi for their support.
\end{acks}

\bibliographystyle{ACM-Reference-Format}
\bibliography{main}


\begin{thebibliography}{29}


\ifx \showCODEN    \undefined \def \showCODEN     #1{\unskip}     \fi
\ifx \showDOI      \undefined \def \showDOI       #1{#1}\fi
\ifx \showISBNx    \undefined \def \showISBNx     #1{\unskip}     \fi
\ifx \showISBNxiii \undefined \def \showISBNxiii  #1{\unskip}     \fi
\ifx \showISSN     \undefined \def \showISSN      #1{\unskip}     \fi
\ifx \showLCCN     \undefined \def \showLCCN      #1{\unskip}     \fi
\ifx \shownote     \undefined \def \shownote      #1{#1}          \fi
\ifx \showarticletitle \undefined \def \showarticletitle #1{#1}   \fi
\ifx \showURL      \undefined \def \showURL       {\relax}        \fi
\providecommand\bibfield[2]{#2}
\providecommand\bibinfo[2]{#2}
\providecommand\natexlab[1]{#1}
\providecommand\showeprint[2][]{arXiv:#2}

\bibitem[Alverti et~al\mbox{.}(2022)]%
        {alverti2022daxvm}
\bibfield{author}{\bibinfo{person}{Chloe Alverti}, \bibinfo{person}{Vasileios Karakostas}, \bibinfo{person}{Nikhita Kunati}, \bibinfo{person}{Georgios Goumas}, {and} \bibinfo{person}{Michael Swift}.} \bibinfo{year}{2022}\natexlab{}.
\newblock \showarticletitle{DaxVM: Stressing the Limits of Memory as a File Interface}. In \bibinfo{booktitle}{\emph{2022 55th IEEE/ACM International Symposium on Microarchitecture (MICRO)}}. IEEE, \bibinfo{pages}{369--387}.
\newblock


\bibitem[Beamer et~al\mbox{.}(2015)]%
        {beamer2015gap}
\bibfield{author}{\bibinfo{person}{Scott Beamer}, \bibinfo{person}{Krste Asanovi{\'c}}, {and} \bibinfo{person}{David Patterson}.} \bibinfo{year}{2015}\natexlab{}.
\newblock \showarticletitle{The GAP benchmark suite}.
\newblock \bibinfo{journal}{\emph{arXiv preprint arXiv:1508.03619}} (\bibinfo{year}{2015}).
\newblock


\bibitem[Busato et~al\mbox{.}(2018)]%
        {busato2018hornet}
\bibfield{author}{\bibinfo{person}{Federico Busato}, \bibinfo{person}{Oded Green}, \bibinfo{person}{Nicola Bombieri}, {and} \bibinfo{person}{David~A Bader}.} \bibinfo{year}{2018}\natexlab{}.
\newblock \showarticletitle{Hornet: An efficient data structure for dynamic sparse graphs and matrices on gpus}. In \bibinfo{booktitle}{\emph{2018 IEEE High Performance extreme Computing Conference (HPEC)}}. IEEE, \bibinfo{pages}{1--7}.
\newblock


\bibitem[Enberg et~al\mbox{.}(2022)]%
        {enberg2022transcending}
\bibfield{author}{\bibinfo{person}{Pekka Enberg}, \bibinfo{person}{Ashwin Rao}, \bibinfo{person}{Jon Crowcroft}, {and} \bibinfo{person}{Sasu Tarkoma}.} \bibinfo{year}{2022}\natexlab{}.
\newblock \showarticletitle{Transcending POSIX: The End of an Era?}
\newblock \bibinfo{journal}{\emph{; login:}} (\bibinfo{year}{2022}).
\newblock


\bibitem[Feng et~al\mbox{.}(2023)]%
        {feng2023tricache}
\bibfield{author}{\bibinfo{person}{Guanyu Feng}, \bibinfo{person}{Huanqi Cao}, \bibinfo{person}{Xiaowei Zhu}, \bibinfo{person}{Bowen Yu}, \bibinfo{person}{Yuanwei Wang}, \bibinfo{person}{Zixuan Ma}, \bibinfo{person}{Shengqi Chen}, {and} \bibinfo{person}{Wenguang Chen}.} \bibinfo{year}{2023}\natexlab{}.
\newblock \showarticletitle{TriCache: A User-Transparent Block Cache Enabling High-Performance Out-of-Core Processing with In-Memory Programs}.
\newblock \bibinfo{journal}{\emph{ACM Transactions on Storage}} \bibinfo{volume}{19}, \bibinfo{number}{2} (\bibinfo{year}{2023}), \bibinfo{pages}{1--30}.
\newblock


\bibitem[Gabert and {\c{C}}ataly{\"u}rek(2021)]%
        {gabert2021pigo}
\bibfield{author}{\bibinfo{person}{Kasimir Gabert} {and} \bibinfo{person}{{\"U}mit~V {\c{C}}ataly{\"u}rek}.} \bibinfo{year}{2021}\natexlab{}.
\newblock \showarticletitle{PIGO: A parallel graph input/output library}. In \bibinfo{booktitle}{\emph{2021 IEEE International Parallel and Distributed Processing Symposium Workshops (IPDPSW)}}. IEEE, \bibinfo{pages}{276--279}.
\newblock


\bibitem[Han et~al\mbox{.}(2013)]%
        {han2013turbograph}
\bibfield{author}{\bibinfo{person}{Wook-Shin Han}, \bibinfo{person}{Sangyeon Lee}, \bibinfo{person}{Kyungyeol Park}, \bibinfo{person}{Jeong-Hoon Lee}, \bibinfo{person}{Min-Soo Kim}, \bibinfo{person}{Jinha Kim}, {and} \bibinfo{person}{Hwanjo Yu}.} \bibinfo{year}{2013}\natexlab{}.
\newblock \showarticletitle{TurboGraph: a fast parallel graph engine handling billion-scale graphs in a single PC}. In \bibinfo{booktitle}{\emph{Proceedings of the 19th ACM SIGKDD international conference on Knowledge discovery and data mining}}. \bibinfo{pages}{77--85}.
\newblock


\bibitem[Imamura and Yoshida(2019)]%
        {imamura2019poster}
\bibfield{author}{\bibinfo{person}{Satoshi Imamura} {and} \bibinfo{person}{Eiji Yoshida}.} \bibinfo{year}{2019}\natexlab{}.
\newblock \showarticletitle{POSTER: AR-MMAP: Write Performance Improvement of Memory-Mapped File}. In \bibinfo{booktitle}{\emph{2019 28th International Conference on Parallel Architectures and Compilation Techniques (PACT)}}. IEEE, \bibinfo{pages}{493--494}.
\newblock


\bibitem[Kolodziej et~al\mbox{.}(2019)]%
        {kolodziej2019suitesparse}
\bibfield{author}{\bibinfo{person}{Scott~P Kolodziej}, \bibinfo{person}{Mohsen Aznaveh}, \bibinfo{person}{Matthew Bullock}, \bibinfo{person}{Jarrett David}, \bibinfo{person}{Timothy~A Davis}, \bibinfo{person}{Matthew Henderson}, \bibinfo{person}{Yifan Hu}, {and} \bibinfo{person}{Read Sandstrom}.} \bibinfo{year}{2019}\natexlab{}.
\newblock \showarticletitle{The suitesparse matrix collection website interface}.
\newblock \bibinfo{journal}{\emph{Journal of Open Source Software}} \bibinfo{volume}{4}, \bibinfo{number}{35} (\bibinfo{year}{2019}), \bibinfo{pages}{1244}.
\newblock


\bibitem[Kyrola et~al\mbox{.}(2012)]%
        {kyrola2012graphchi}
\bibfield{author}{\bibinfo{person}{Aapo Kyrola}, \bibinfo{person}{Guy Blelloch}, {and} \bibinfo{person}{Carlos Guestrin}.} \bibinfo{year}{2012}\natexlab{}.
\newblock \showarticletitle{$\{$GraphChi$\}$:$\{$Large-Scale$\}$ graph computation on just a $\{$PC$\}$}. In \bibinfo{booktitle}{\emph{10th USENIX symposium on operating systems design and implementation (OSDI 12)}}. \bibinfo{pages}{31--46}.
\newblock


\bibitem[Leis et~al\mbox{.}(2023)]%
        {leis2023virtual}
\bibfield{author}{\bibinfo{person}{Viktor Leis}, \bibinfo{person}{Adnan Alhomssi}, \bibinfo{person}{Tobias Ziegler}, \bibinfo{person}{Yannick Loeck}, {and} \bibinfo{person}{Christian Dietrich}.} \bibinfo{year}{2023}\natexlab{}.
\newblock \showarticletitle{Virtual-Memory Assisted Buffer Management}.
\newblock \bibinfo{journal}{\emph{Proceedings of the ACM on Management of Data}} \bibinfo{volume}{1}, \bibinfo{number}{1} (\bibinfo{year}{2023}), \bibinfo{pages}{1--25}.
\newblock


\bibitem[Li et~al\mbox{.}(2019)]%
        {li2019userland}
\bibfield{author}{\bibinfo{person}{Feng Li}, \bibinfo{person}{Daniel~G Waddington}, {and} \bibinfo{person}{Fengguang Song}.} \bibinfo{year}{2019}\natexlab{}.
\newblock \showarticletitle{Userland CO-PAGER: boosting data-intensive applications with non-volatile memory, userspace paging}. In \bibinfo{booktitle}{\emph{Proceedings of the 3rd International Conference on High Performance Compilation, Computing and Communications}}. \bibinfo{pages}{78--83}.
\newblock


\bibitem[Lin et~al\mbox{.}(2014)]%
        {lin2014mmap}
\bibfield{author}{\bibinfo{person}{Zhiyuan Lin}, \bibinfo{person}{Minsuk Kahng}, \bibinfo{person}{Kaeser~Md Sabrin}, \bibinfo{person}{Duen Horng~Polo Chau}, \bibinfo{person}{Ho Lee}, {and} \bibinfo{person}{U Kang}.} \bibinfo{year}{2014}\natexlab{}.
\newblock \showarticletitle{Mmap: Fast billion-scale graph computation on a pc via memory mapping}. In \bibinfo{booktitle}{\emph{2014 IEEE International Conference on Big Data (Big Data)}}. IEEE, \bibinfo{pages}{159--164}.
\newblock


\bibitem[Malliotakis et~al\mbox{.}(2021)]%
        {malliotakis2021hugemap}
\bibfield{author}{\bibinfo{person}{Ioannis Malliotakis}, \bibinfo{person}{Anastasios Papagiannis}, \bibinfo{person}{Manolis Marazakis}, {and} \bibinfo{person}{Angelos Bilas}.} \bibinfo{year}{2021}\natexlab{}.
\newblock \showarticletitle{HugeMap: Optimizing Memory-Mapped I/O with Huge Pages for Fast Storage}. In \bibinfo{booktitle}{\emph{Euro-Par 2020: Parallel Processing Workshops: Euro-Par 2020 International Workshops, Warsaw, Poland, August 24--25, 2020, Revised Selected Papers 26}}. Springer, \bibinfo{pages}{344--355}.
\newblock


\bibitem[Najeebullah et~al\mbox{.}(2014)]%
        {najeebullah2014bishard}
\bibfield{author}{\bibinfo{person}{Kamran Najeebullah}, \bibinfo{person}{Kifayat~Ullah Khan}, \bibinfo{person}{Waqas Nawaz}, {and} \bibinfo{person}{Young-Koo Lee}.} \bibinfo{year}{2014}\natexlab{}.
\newblock \showarticletitle{Bishard parallel processor: A disk-based processing engine for billion-scale graphs}.
\newblock \bibinfo{journal}{\emph{International Journal of Multimedia and Ubiquitous Engineering}} \bibinfo{volume}{9}, \bibinfo{number}{2} (\bibinfo{year}{2014}), \bibinfo{pages}{199--212}.
\newblock


\bibitem[Nguyen et~al\mbox{.}(2013)]%
        {nguyen2013lightweight}
\bibfield{author}{\bibinfo{person}{Donald Nguyen}, \bibinfo{person}{Andrew Lenharth}, {and} \bibinfo{person}{Keshav Pingali}.} \bibinfo{year}{2013}\natexlab{}.
\newblock \showarticletitle{A lightweight infrastructure for graph analytics}. In \bibinfo{booktitle}{\emph{Proceedings of the twenty-fourth ACM symposium on operating systems principles}}. \bibinfo{pages}{456--471}.
\newblock


\bibitem[Papagiannis et~al\mbox{.}(2021)]%
        {papagiannis2021memory}
\bibfield{author}{\bibinfo{person}{Anastasios Papagiannis}, \bibinfo{person}{Manolis Marazakis}, {and} \bibinfo{person}{Angelos Bilas}.} \bibinfo{year}{2021}\natexlab{}.
\newblock \showarticletitle{Memory-mapped I/O on steroids}. In \bibinfo{booktitle}{\emph{Proceedings of the Sixteenth European Conference on Computer Systems}}. \bibinfo{pages}{277--293}.
\newblock


\bibitem[Papagiannis et~al\mbox{.}(2020)]%
        {papagiannis2020optimizing}
\bibfield{author}{\bibinfo{person}{Anastasios Papagiannis}, \bibinfo{person}{Giorgos Xanthakis}, \bibinfo{person}{Giorgos Saloustros}, \bibinfo{person}{Manolis Marazakis}, {and} \bibinfo{person}{Angelos Bilas}.} \bibinfo{year}{2020}\natexlab{}.
\newblock \showarticletitle{Optimizing memory-mapped $\{$I/O$\}$ for fast storage devices}. In \bibinfo{booktitle}{\emph{2020 USENIX Annual Technical Conference (USENIX ATC 20)}}. \bibinfo{pages}{813--827}.
\newblock


\bibitem[Roy et~al\mbox{.}(2013)]%
        {roy2013x}
\bibfield{author}{\bibinfo{person}{Amitabha Roy}, \bibinfo{person}{Ivo Mihailovic}, {and} \bibinfo{person}{Willy Zwaenepoel}.} \bibinfo{year}{2013}\natexlab{}.
\newblock \showarticletitle{X-stream: Edge-centric graph processing using streaming partitions}. In \bibinfo{booktitle}{\emph{Proceedings of the Twenty-Fourth ACM Symposium on Operating Systems Principles}}. \bibinfo{pages}{472--488}.
\newblock


\bibitem[Shun and Blelloch(2013)]%
        {shun2013ligra}
\bibfield{author}{\bibinfo{person}{Julian Shun} {and} \bibinfo{person}{Guy~E Blelloch}.} \bibinfo{year}{2013}\natexlab{}.
\newblock \showarticletitle{Ligra: a lightweight graph processing framework for shared memory}. In \bibinfo{booktitle}{\emph{Proceedings of the 18th ACM SIGPLAN symposium on Principles and practice of parallel programming}}. \bibinfo{pages}{135--146}.
\newblock


\bibitem[Song et~al\mbox{.}(2016)]%
        {song2016efficient}
\bibfield{author}{\bibinfo{person}{Nae~Young Song}, \bibinfo{person}{Yongseok Son}, \bibinfo{person}{Hyuck Han}, {and} \bibinfo{person}{Heon~Young Yeom}.} \bibinfo{year}{2016}\natexlab{}.
\newblock \showarticletitle{Efficient memory-mapped I/O on fast storage device}.
\newblock \bibinfo{journal}{\emph{ACM Transactions on Storage (TOS)}} \bibinfo{volume}{12}, \bibinfo{number}{4} (\bibinfo{year}{2016}), \bibinfo{pages}{1--27}.
\newblock


\bibitem[Song et~al\mbox{.}(2012)]%
        {song2012low}
\bibfield{author}{\bibinfo{person}{Nae~Young Song}, \bibinfo{person}{Young~Jin Yu}, \bibinfo{person}{Woong Shin}, \bibinfo{person}{Hyeonsang Eom}, {and} \bibinfo{person}{Heon~Young Yeom}.} \bibinfo{year}{2012}\natexlab{}.
\newblock \showarticletitle{Low-latency memory-mapped i/o for data-intensive applications on fast storage devices}. In \bibinfo{booktitle}{\emph{2012 SC Companion: High Performance Computing, Networking Storage and Analysis}}. IEEE, \bibinfo{pages}{766--770}.
\newblock


\bibitem[Staudt et~al\mbox{.}(2016)]%
        {staudt2016networkit}
\bibfield{author}{\bibinfo{person}{Christian~L Staudt}, \bibinfo{person}{Aleksejs Sazonovs}, {and} \bibinfo{person}{Henning Meyerhenke}.} \bibinfo{year}{2016}\natexlab{}.
\newblock \showarticletitle{NetworKit: A tool suite for large-scale complex network analysis}.
\newblock \bibinfo{journal}{\emph{Network Science}} \bibinfo{volume}{4}, \bibinfo{number}{4} (\bibinfo{year}{2016}), \bibinfo{pages}{508--530}.
\newblock


\bibitem[Van~Essen et~al\mbox{.}(2015)]%
        {van2015di}
\bibfield{author}{\bibinfo{person}{Brian Van~Essen}, \bibinfo{person}{Henry Hsieh}, \bibinfo{person}{Sasha Ames}, \bibinfo{person}{Roger Pearce}, {and} \bibinfo{person}{Maya Gokhale}.} \bibinfo{year}{2015}\natexlab{}.
\newblock \showarticletitle{DI-MMAP—a scalable memory-map runtime for out-of-core data-intensive applications}.
\newblock \bibinfo{journal}{\emph{Cluster Computing}}  \bibinfo{volume}{18} (\bibinfo{year}{2015}), \bibinfo{pages}{15--28}.
\newblock


\bibitem[Wang et~al\mbox{.}(2021)]%
        {wang2021scaleg}
\bibfield{author}{\bibinfo{person}{Xubo Wang}, \bibinfo{person}{Dong Wen}, \bibinfo{person}{Lu Qin}, \bibinfo{person}{Lijun Chang}, \bibinfo{person}{Ying Zhang}, {and} \bibinfo{person}{Wenjie Zhang}.} \bibinfo{year}{2021}\natexlab{}.
\newblock \showarticletitle{Scaleg: A distributed disk-based system for vertex-centric graph processing}.
\newblock \bibinfo{journal}{\emph{IEEE Transactions on Knowledge and Data Engineering}} \bibinfo{volume}{35}, \bibinfo{number}{2} (\bibinfo{year}{2021}), \bibinfo{pages}{2019--2033}.
\newblock


\bibitem[Wang et~al\mbox{.}(2016)]%
        {wang2016gunrock}
\bibfield{author}{\bibinfo{person}{Yangzihao Wang}, \bibinfo{person}{Andrew Davidson}, \bibinfo{person}{Yuechao Pan}, \bibinfo{person}{Yuduo Wu}, \bibinfo{person}{Andy Riffel}, {and} \bibinfo{person}{John~D Owens}.} \bibinfo{year}{2016}\natexlab{}.
\newblock \showarticletitle{Gunrock: A high-performance graph processing library on the GPU}. In \bibinfo{booktitle}{\emph{Proceedings of the 21st ACM SIGPLAN symposium on principles and practice of parallel programming}}. \bibinfo{pages}{1--12}.
\newblock


\bibitem[Yang et~al\mbox{.}(2022)]%
        {yang2022graphblast}
\bibfield{author}{\bibinfo{person}{Carl Yang}, \bibinfo{person}{Ayd{\i}n Bulu{\c{c}}}, {and} \bibinfo{person}{John~D Owens}.} \bibinfo{year}{2022}\natexlab{}.
\newblock \showarticletitle{GraphBLAST: A high-performance linear algebra-based graph framework on the GPU}.
\newblock \bibinfo{journal}{\emph{ACM Transactions on Mathematical Software (TOMS)}} \bibinfo{volume}{48}, \bibinfo{number}{1} (\bibinfo{year}{2022}), \bibinfo{pages}{1--51}.
\newblock


\bibitem[Yoshimura et~al\mbox{.}(2019)]%
        {yoshimura2019evfs}
\bibfield{author}{\bibinfo{person}{Takeshi Yoshimura}, \bibinfo{person}{Tatsuhiro Chiba}, {and} \bibinfo{person}{Hiroshi Horii}.} \bibinfo{year}{2019}\natexlab{}.
\newblock \showarticletitle{$\{$EvFS$\}$: User-level,$\{$Event-Driven$\}$ File System for $\{$Non-Volatile$\}$ Memory}. In \bibinfo{booktitle}{\emph{11th USENIX Workshop on Hot Topics in Storage and File Systems (HotStorage 19)}}.
\newblock


\bibitem[Zheng et~al\mbox{.}(2015)]%
        {zheng2015flashgraph}
\bibfield{author}{\bibinfo{person}{Da Zheng}, \bibinfo{person}{Disa Mhembere}, \bibinfo{person}{Randal Burns}, \bibinfo{person}{Joshua Vogelstein}, \bibinfo{person}{Carey~E Priebe}, {and} \bibinfo{person}{Alexander~S Szalay}.} \bibinfo{year}{2015}\natexlab{}.
\newblock \showarticletitle{$\{$FlashGraph$\}$: Processing $\{$Billion-Node$\}$ graphs on an array of commodity $\{$SSDs$\}$}. In \bibinfo{booktitle}{\emph{13th USENIX Conference on File and Storage Technologies (FAST 15)}}. \bibinfo{pages}{45--58}.
\newblock


\end{thebibliography}
\end{document}